# Comment on "Experimental Studies of Os⁻: Observation of a Bound-Bound Electric Dipole Transition in an Atomic Negative Ion"

Zineb Felfli[1], Filmon Kiros[1], Kelvin Suggs[2] and Alfred Z. Msezane[1]
[1]Department of Physics and CTSPS, Clark Atlanta University, Atlanta, Georgia 30314, USA
[2]Department of Chemistry, Clark Atlanta University, Atlanta, Georgia 30314, USA

Bilodeau and Haugen [1], using Infrared laser photodetachment spectroscopy, measured the binding energies (BEs) of the ground state ($^4F^e_{9/2}$) and the excited state ($^4F^e_{7/2}$) of the Os⁻ ion to be 1.07780(12) eV and 0.553(3) eV, respectively. These values are consistent with those calculated using Relativistic Configuration Interaction (RCI) calculations [2]. Here we have calculated the BEs for the ground state and the two excited states of the Os⁻ ion using our recent complex angular momentum (CAM) methodology [3] and obtained the BEs of 1.910, 1.230 and 0.224 eV, respectively (see Figure). We conclude that: 1) the measured value of 1.07780(12) eV corresponds to an excited state of Os⁻ and not to the EA of Os and 2) the EA of Os is 1.910 eV.

Laser cooling of atomic anions has been suggested for producing ultra-cold antihydrogens [4] whose important scientific interest [5] includes the direct measurement of gravitational acceleration of antimatter [6]. Hence the great need for accurate data for the Os⁻ ion. Atomic negative ions are also important in catalysis [7]. Preliminary results using the Os⁻ ion show dramatic improvement on the catalysis of $H_2O$ to $H_2O_2$ over recent results obtained using Au⁻ and Pd⁻ ions [8] and in the oxidation of $CH_4$ to methanol [9] using the Au⁻ ion.

In calculations involving negative ions, the diffuse nature of the orbitals necessitates using an extensive partial wave expansion. This approach yields results that are often riddled with uncertainties and lack definitiveness for complex systems [10]. In the present case we consider the formation of the atomic Os⁻ negative ion as a Regge resonance through the elastic collision of a slow electron with atomic Os and extract the BEs from the electron elastic scattering total cross sections (TCSs). Embedded in our Regge-pole methodology [3] are the crucial electron-electron correlations and the vital core polarization interactions, which are responsible for the existence and stability of typical negative ions. Importantly, accounting for relativistic effects does not necessarily guarantee reliable results, if the crucial effects are inadequately accounted for [10].

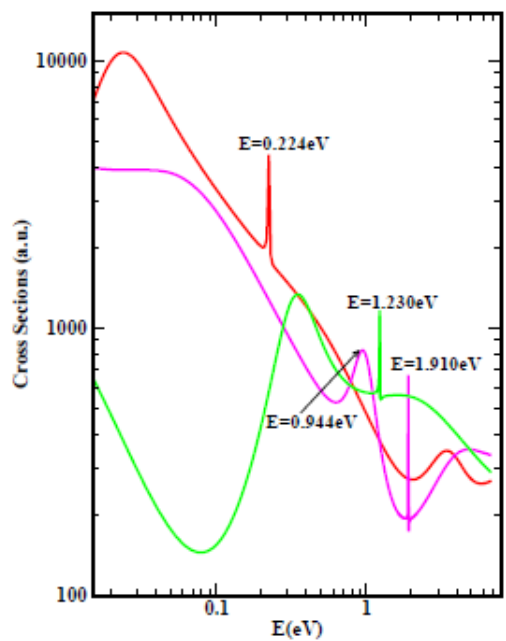

Our results, seen from figure, demonstrate that the TCS for Os is characterized by a complex resonance structure, ideal for catalysis, making it difficult to execute the Wigner threshold law in describing the threshold detachment behavior of complex atoms and extracting the reliable attendant EA.

___________________